\documentclass[%
 reprint,
superscriptaddress,
 amsmath,amssymb,
 aps,
floatfix,
float,
placeins
]{revtex4-2}
\usepackage{xcolor}
\usepackage[normalem]{ulem}
\usepackage{graphicx}
\usepackage{dcolumn}
\usepackage{bm}

\begin{document}

\preprint{APS/123-QED}

\title{Spin pumping driven by magnon-polaritons in a ferromagnet-coplanar superconducting resonator hybrid system}

\author{Dinesh Wagle}
\affiliation{Department of Physics and Astronomy, University of Delaware, Newark, DE 19716, USA}
\author{Yi Li}
\affiliation{Materials Science Division, Argonne National Laboratory,  Lemont, IL 60439, USA}
\author{Anish Rai}
\affiliation{Department of Physics and Astronomy, University of Delaware, Newark, DE 19716, USA}
\author{Tomas Polakovic}
\affiliation{Physics Division, Argonne National Laboratory,  Lemont, IL 60439, USA}
\author{Valentine Novosad}
\affiliation{Materials Science Division, Argonne National Laboratory,  Lemont, IL 60439, USA}
\author{M. Benjamin Jungfleisch}
\email{mbj@udel.edu}
\affiliation{Department of Physics and Astronomy, University of Delaware, Newark, DE 19716, USA}

\date{\today}

\begin{abstract}
We demonstrate spin pumping driven by a strongly coupled magnon-photon system using a ferromagnet-coplanar superconducting resonator hybrid system at 1.4 K. Electrical readout via the inverse spin-Hall effect reveals characteristic coupling features, including mode splitting and linewidth broadening, demonstrating the electrical detection of strongly coupled microwave photons and magnons. The magnon-photon coupling strength obtained by combined spin pumping and inverse spin-Hall effect measurements is compared to microwave transmission experiments. Furthermore, microwave power-dependent measurements reveal a decrease in the coupling strength with increasing microwave power alongside the onset of nonlinearities of the superconducting resonator above a critical microwave power threshold.
\end{abstract}

\maketitle

\section{\label{sec:intro} Introduction}

A hybrid quantum system combines two or more systems with complementary functionalities to leverage their individual strengths and overcome specific limitations. These systems often integrate various types of quantum technologies, such as superconducting qubits, spins, trapped ions, photons, phonons, and quantum dots, to achieve more robust and versatile capabilities in quantum computation, communication, transduction, and sensing~\cite{tabuchi_sci_15, Zhang_PhysRevLett_2014, Sutherland_PRA_2021, DeCrescent_PhysRevApplied_2022}. The efficiency of energy transfer between these systems depends on the strength of coupling. However, on-demand tuning of the coupling properties, i.e., modulating the coupling rates and/or the dissipation rates of the subsystems in an on-chip integrated device remains a challenge.

In magnonics, magnons -- the quanta of collective spin excitations -- serve as fundamental information carriers for next-generation wave-based computing and communication technologies~\cite{Lachance-Quirion_APE_2019}. Their charge-free nature makes them an attractive platform for ultralow-power devices and circuits, as they exhibit lower energy dissipation typically associated with Ohmic losses in conventional charge-current-based systems~\cite{Chumak_NatPhys_2015}. Magnons can interact strongly with microwave photons~\cite{Artman_PhysRev._1953, Soykal_PRL_2010}, phonons~\cite{Zhang_SciAdv_16}, and other magnons~\cite{Chen_PRl_2018, Klingler_2016, wagle_arxiv_2024}, giving rise to dynamic hybrid systems. In particular, the interaction between magnons and microwave photons -- mediated by the dipolar (Zeeman) interaction -- leads to the formation of magnon-polaritons, which exhibit a wide range of rich and tunable physical phenomena. These hybrid excitations have been extensively studied using microwave spectroscopy techniques in both 3D cavities~\cite{Tabuchi_PhysRevLett_2014, Zhang_PhysRevLett_2014} and planar resonators~\cite{Huebl_PhysRevLett_2013, Bhoi_JAP_2014, Li_PRL_2019}, as well as through optical detection methods~\cite{Osada_2018, Hisatomi_2016, Klingler_2016, Kaffash_QST_2023} and electrical detection using spin pumping and inverse spin Hall effect~\cite{Bai_PhysRevLett_2015, Xu_PhysRevB_2020}. The avoided crossing between the photon and magnon modes is the characteristic fingerprint of (strong) magnon-photon coupling.

The strength of this coupling relative to the losses in the hybrid system is quantified by the cooperativity $C$, given by
\begin{equation}
C = \frac{g^2}{\kappa_p \kappa_m},
\label{eq:strong}
\end{equation}
where $g$ is the coupling strength defined by the avoided crossing gap, and $\kappa_p$ and $\kappa_m$ represent the photon and magnon loss rates, respectively. A high cooperativity indicates strong coherent interaction between the two subsystems. This can be achieved either by increasing the coupling strength~\cite{Zhang_PhysRevLett_2014, Wagle_JPMaterials_2024} or by minimizing losses in the system -- achievable through the use of high-Q magnonic and/or photonic resonators~\cite{Li_PRL_2019, Xu_AdvSci_2024}.

The electrical readout of magnon-polaritons in an high-cooperativity system at low temperature would be a significant advance as it could provide a pathway for electrically detecting light/matter (microwave/magnet) interaction at the quantum level. However, electrical readout of information about magnons strongly coupled to microwave photons by spin pumping is detrimental to the strong coupling behavior because of an enhanced damping induced by spin pumping, i.e., $\kappa_m$ in Eq.~(\ref{eq:strong}).

Superconducting (SC) circuits with low-loss operation and easy on-chip integrability offer a potential solution to detecting magnons strongly coupled to microwave photons by spin pumping at low temperatures. They provide several key advantages, including compatibility with existing semiconductor technologies, low error rates, and scalability. Superconducting resonators, which are crucial components of such circuits, feature smaller mode volumes and higher quality factors~\cite{Huebl_PhysRevLett_2013, Li_PRL_2019, Igor_SciAdv_2021, Ghirri_PhysRevApplied_2023, Morris_SciRep_2017, Li_PhysRevLett_2022}. These resonators are commonly used in astronomy for ultra-low-noise measurements~\cite{Wandui_JAP_2020} and in high-energy astrophysics for photon-counting experiments~\cite{Ulbricht_APL_2015}. They can be coupled to tunnel junctions to create qubits~\cite{Martinis_QIP_2009} and embedded spin systems to create memory elements for computing applications~\cite{MORTON_JMR_2018}. Additionally, they are widely used in hybrid magnonic systems to achieve high cooperativity~\cite{ Huebl_PhysRevLett_2013, Li_PRL_2019, McKenzie_PhysRevB_2019, Mandal_APL_2020, Baity_APL_2021, Haygood_PRA_2021}.

\begin{figure}[t]
\centering
\includegraphics[width=1\columnwidth]{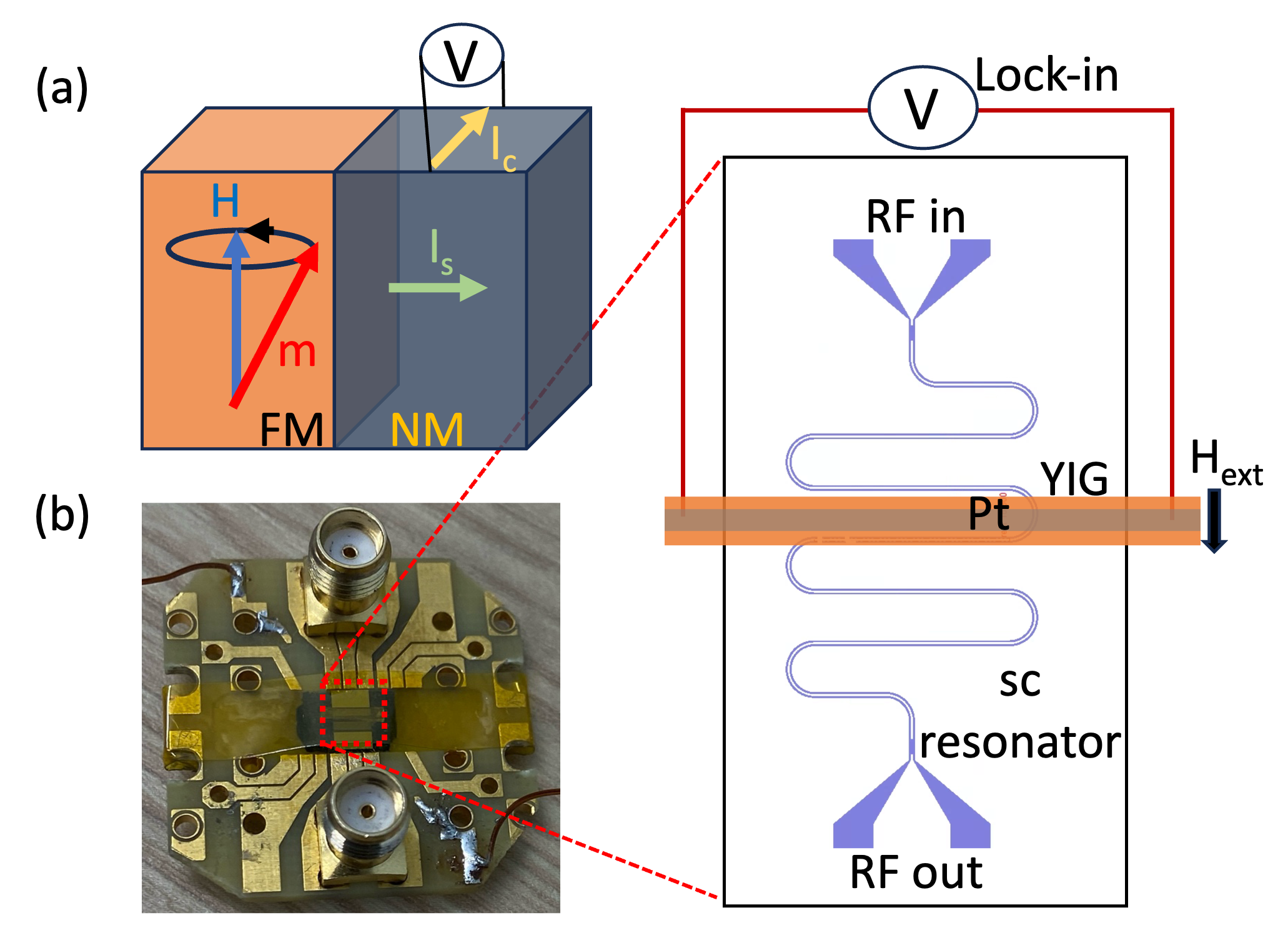}
\caption{(a) Schematic illustration of spin pumping process. A precessing magnetization in the FM layer drives a spin current $I_s$ across the interface which is converted to charge current $I_c$ via the inverse spin Hall effect. (b) Experimental setup: the YIG/Pt bilayer is flip-chipped on top of SC resonator, while a rf current is passed through the resonator and an external magnetic field is applied perpendicular to the Pt stripe. The spin pumping voltage is measured across the two ends of the Pt stripe.
}
\label{Fig1}
\end{figure}

In this article, we address two key questions: Can magnon-polaritons be electrically detected in a nanometer-thin sample at cryogenic temperature? How can the properties of a magnon-photon coupled system be tuned on demand by microwave power? We demonstrate the electrical detection of coherent magnon-photon coupling by spin pumping and inverse spin-Hall effect at 1.4 K [see Fig.~\ref{Fig1}(a)]. Our hybrid system comprises a high quality superconducting resonator and an yttrium iron garnet (YIG)/ platinum (Pt) bilayer thin film. The electrical measurements are compared to microwave transmission experiments.
In accordance with theory, we find that the linewidth of the magnon mode increases when it approaches the coupling region. Furthermore, we reveal that the coupling strength decreases as the applied microwave power increases. This behavior is attributed to the onset of nonlinearities in the superconducting resonator, which appears above a critical microwave power threshold.  
Our results provide a pathway towards the electrical detection of the magnonic properties of a magnet strongly coupled to microwaves in the quantum regime.

\section{\label{sec:exp} Experimental details}

Figure~\ref{Fig1} shows the experimental setup along with an illustration of the spin pumping process, where a pure spin current is generated by a precessing ferromagnet.  
A 200-nm-thick commercial YIG film, grown on both sides of a 500 $\mu$m-thick gadolinium gallium garnet (GGG) substrate via liquid phase epitaxy, was used in this study. The film, characterized by a damping $\alpha$ of $8.5\pm1.0\times 10^{-4}$ , was diced into a 5 mm $\times$ 1 mm rectangular sample. A 3 nm-thick Pt layer was then deposited on top of the YIG sample by DC magnetron sputtering at a base pressure of 3$\times$10$^{-8}$ Torr and in Ar atmosphere at 3 mTorr using a shadow mask. Following the Pt deposition, the damping increased to $\alpha_{\mathrm{eff}}$ = $9.9\pm1.0\times 10^{-4}$, leading to a spin-mixing conductance of $g^{\uparrow\downarrow}$ = $2.8\pm0.3\times10^{18}$ m$^{-2}$, consistent with the values reported in the literature~\cite{Jungfleisch_PhysRevB_2015}.

Superconducting resonators were fabricated from 200-nm-thick NbN films using standard photolithography followed by reactive ion etching. Prior to film deposition, the undoped Si substrates were cleaned using a low-energy argon ion beam to remove surface contaminants. NbN films were then deposited via DC magnetron sputtering at room temperature under ultra-high vacuum of less than 5$\times 10^{-8}$ Torr and Ar gas pressure of 2 mTorr, with a deposition rate of approximately 1~\text{\AA}/s~\cite{Polakovic_APLM_2018}. The unloaded resonator exhibited a resonance frequency of approximately 6.03 GHz at 1.4 K, which redshifted to $\sim$5.8 GHz upon loading with the YIG/Pt bilayer. The unloaded resonator had a quality factor (Q-factor) of 1210, as shown in the Supplemental Material (SM Fig.~1); the Q-factor of the resonator when loaded with the YIG/Pt bilayer was reduced to 105.

A photograph of the sample holder and the sample orientation is shown in Fig.~\ref{Fig1}(b). The SC resonator was wire-bonded to a PC board to supply a microwave current. The YIG/Pt bilayer was flip-chipped onto the resonator, with two wires connected to the two ends of the Pt stripe via silver epoxy. The entire setup was placed in a cryogenic bath, and an external magnetic field was applied perpendicular to the length of the sample, as illustrated in Fig.~\ref{Fig1}(b). All measurements were performed at 1.4 K, well below the superconducting transition temperature of the NbN resonator (T$_c$ = 14 K~\cite{Li_PRL_2019, Polakovic_APLM_2018}).
The DC component of the pumped spin current was converted into a DC charge current via the inverse spin Hall effect in the Pt layer. This spin pumping voltage was detected by measuring the voltage across the two ends using a lock-in technique, as illustrated in the inset of Fig.~\ref{Fig1}(b). In addition to electrical detection, microwave transmission measurements were performed under identical cryogenic and magnetic field conditions using a vector network analyzer (VNA)-FMR technique, not shown in Fig.~\ref{Fig1}.

\begin{figure}[t]
\centering
\includegraphics[width=1.0\columnwidth]{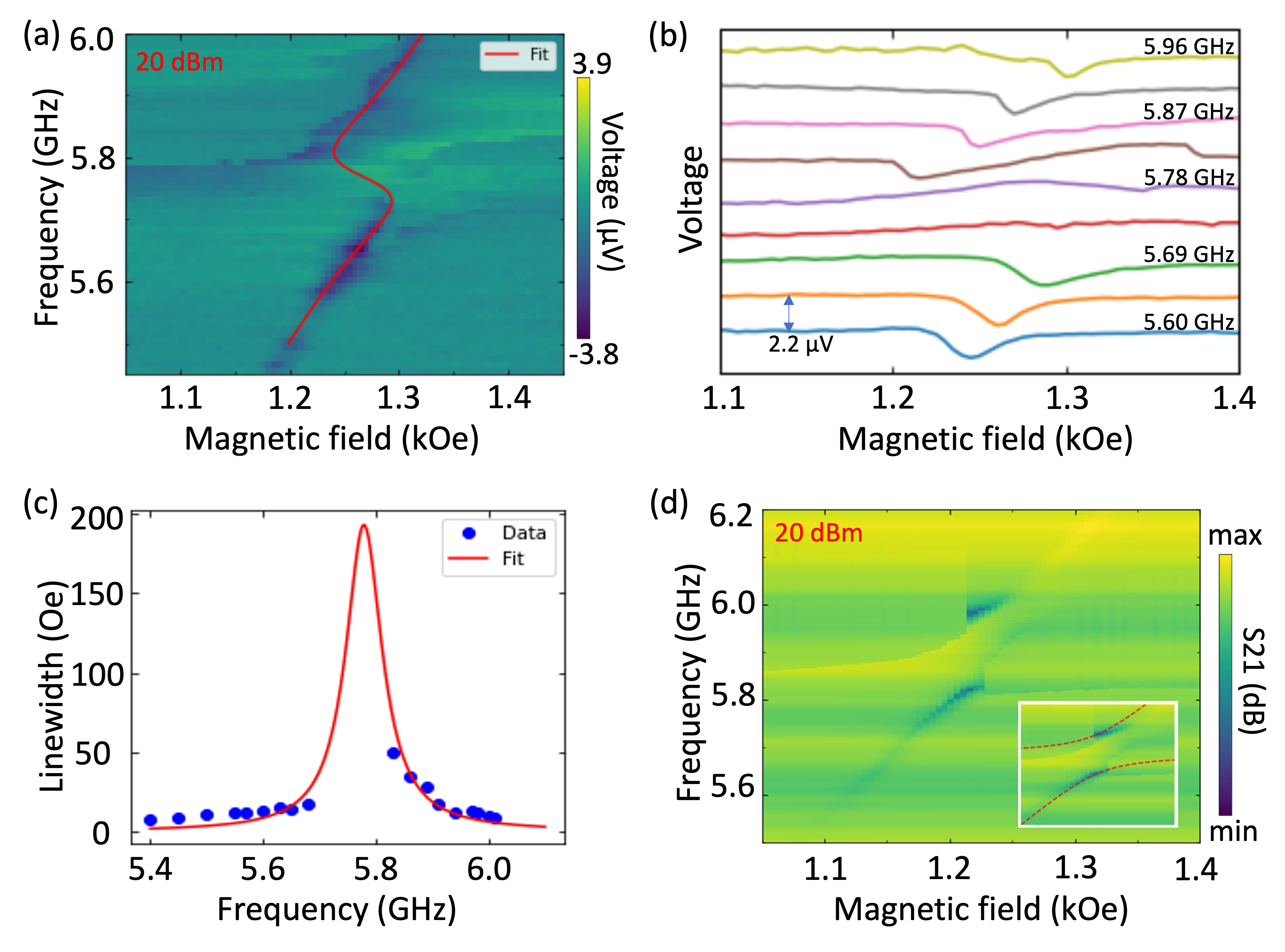}
\caption{(a) Spin pumping voltage spectra as a function of microwave frequency and magnetic field. The red curve is the fitting to the data using real part of Eq.~(\ref{fit_eqn}). (b) Individual voltage plots against magnetic field at fixed microwave frequency corresponding; same data as in (a). (c) The variation of magnon linewidth $\Delta H$ with frequency $f$. The red curve is the fit to the imaginary part of Eq.~(\ref{fit_eqn}). (d) Corresponding VNA spectra. The red curve in the inset represents a fit to the real part of the coupled harmonic oscillator equation, as presented in the Supplemental Material [Eq.~(SM1)].
}
\label{Fig2}
\end{figure}

\section{Results and Discussion}
The spin-pumping measurements were carried out by fixing the microwave frequency and sweeping the magnetic field. In order to obtain an appreciable electrical signal, a comparably large microwave input power was used for the spin pumping experiments. The measured spin-pumping voltage as a function of frequency and the externally applied field for a $\sim$ 5.8 GHz SC resonator at +20~dBm microwave power is shown in Fig.~\ref{Fig2}(a), where the false color represents the spin-pumping voltage. We observe a Kittel-like mode that increases in frequency as the field is increased. However, we observe a characteristic mode hybridization at approximately $1.25$~kOe when the magnon mode and the resonator mode are degenerate. The data unambiguously demonstrates the successful electrical detection of magnon polaritons at cryogenic temperature. To further corroborate our results, we fit the data using the dispersion relation obtained by solving the eigenvalue equation given in Ref.~\cite{Bai_PhysRevLett_2015}:

\begin{small}
\begin{multline}
H = \frac{1}{2} \Bigg[ -4\pi M_{\text{eff}} \\
\pm \sqrt{(4\pi M_{\text{eff}})^2 
+ \left( 
\frac{4\left(f^2 - f_c f (1 - i\beta) - g^2\right)}
{ \left(f(1 - i\alpha) - f_c(1 - i\alpha)(1 - i\beta)\right)\gamma }
\right)^2 } 
\Bigg],
\label{fit_eqn}
\end{multline}
\end{small}

where $\alpha$ is the magnon damping parameter and $\beta$ is the resonator intrinsic loss rate. The real and imaginary parts of the equation represent the resonance field position and the linewidth of the coupled modes at a given microwave frequency, respectively. The fit to the real part of Eq.~(\ref{fit_eqn}) is shown as a red line in Fig.~\ref{Fig2}(a). From the fit, we obtain an effective magnetization 
4$\pi$M$_{eff}$ = 2.03 kOe, 
which is higher than the saturation magnetization value of YIG at room temperature ($\sim$1.75 $\mathrm{kOe}$) but agrees with the literature values for thin YIG films at low temperatures~\cite{Beaulieu_IEEE_2018}. Further, fitting yields the gyromagnetic ratio
$\gamma$ = 2.83 GHz/kOe, which is in close agreement with the literature~\cite{Hauser_SciRep_2016, Serga_JPD_2010}, $\alpha=9.9\pm 3\times 10^{-4}$, and $\beta=5.5\pm 0.6\times 10^{-3}$. Finally, a coupling strength $g$ = 105$\pm$5 MHz is extracted, which is higher than the value reported for comparable YIG film thicknesses at room temperature~\cite{Kaffash_QST_2023}.

The individual line plots of the spin pumping data obtained by sweeping the external magnetic field at fixed microwave frequency are shown in Fig.~\ref{Fig2}(b). The lineshapes are approximately symmetric far away from the coupling region and become asymmetric as the coupling region is approached. This can be understood as follows:
The spin pumping voltage ($V_{SP}$) is related to magnetization ($m$) as $V_{\text{SP}} \propto |m|^2 \propto {(\Delta H)^2}/{\big((H - H_r)^2 + (\Delta H)^2}\big)$~\cite{Bai_PhysRevLett_2013, Bai_PhysRevLett_2015}, which is Lorentzian. The magnetization dynamics far from the coupling region are dominated by intrinsic damping mechanisms, and the microwave resonator has negligible influence. As a result, the lineshape away from the coupling region is Lorentzian.
However, near the coupling region, the magnon and photon modes hybridize, causing a
phase correlation between the microwave current and the magnetization precession to distort and, hence, the lineshape becomes asymmetric~\cite{Bai_PhysRevLett_2015}. 

At the resonance frequency of the SC resonator of $\sim$ 5.8~GHz, a vanishingly small signal is observed. The extracted magnon linewidth $\Delta H$ from the spin pumping data is plotted as a function of microwave frequency $f$ in Fig.~\ref{Fig2}(c). A noticeable increase in linewidth is found near the coupling region.
The red line represents a fit to the imaginary part of Eq.~(\ref{fit_eqn}) simultaneously  fitted with the real part shown in Fig.~\ref{Fig2}(a).
This increase in $\Delta H$ is due to a coherent coupling of the magnon mode with the microwave field~\cite{Bai_PhysRevLett_2015}.

The corresponding microwave spectroscopy data measured under identical cryogenic and magnetic field conditions using a VNA and at a microwave power of +20 dBm is shown in Fig.~\ref{Fig2}(d) where the false color represents the transmission parameter (S${21}$). The observed avoided crossing is due to the hybridization of magnon and photon modes. 
The spectrum is fitted to the real part of the coupled harmonic oscillator model, solved for the complex frequency $f$ at a fixed magnetic field $H$, as described in the Supplemental Material (SM) [Eq.~(SM1)]. The resulting avoided crossing fit is shown as a red dotted line in the inset of Fig.~\ref{Fig2}(d). The fit yields $4\pi M_{eff} = 2.02\pm0.10$ kOe, $\alpha = 1.0\pm0.3\times10^{-3}$ and $\beta = 4.0\pm1.1\times10^{-3}$, in close agreement with the parameters extracted from the spin-pumping measurements. Additionally, the coupling strength is extracted to be $g = 90\pm4$ MHz, which is approximately $14\%$ lower than the value obtained via spin pumping.

Another interesting observation is that the resonator mode at +20 dBm microwave power becomes fainter as we move away from the coupling region, where magnon and resonator mode are degenerate. This is a highly unusual behavior as the resonator mode is typically most intense as nearly all power is transmitted far away from the coupling region. Despite this anomaly, the coupling region can be clearly identified, and the data remain amenable to fitting, allowing us to extract important parameters from the spectrum. 

\begin{figure}[t]
\centering
\includegraphics[width=1.0\columnwidth]{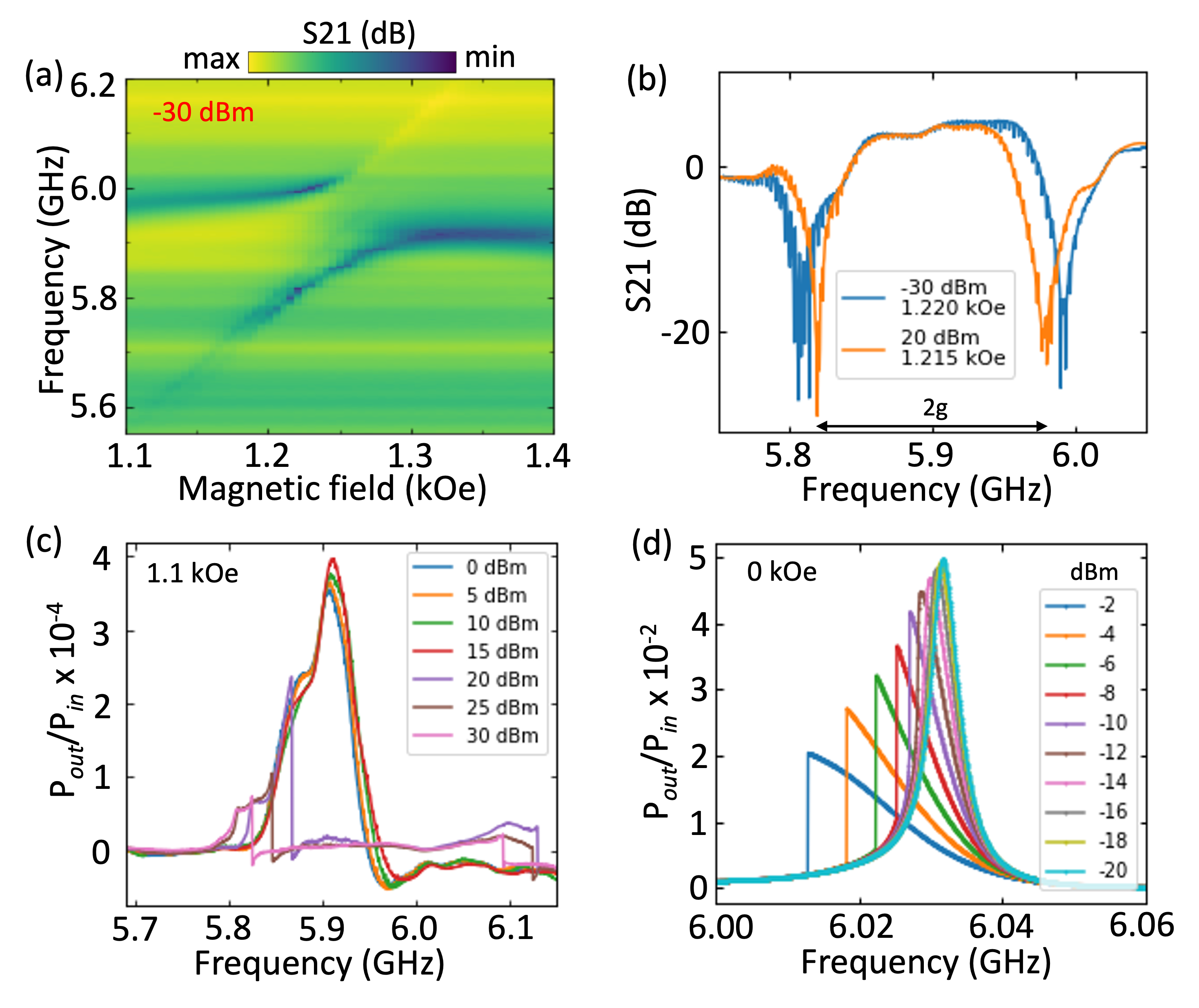}
\caption{(a) A typical avoided level crossing spectrum obtained using VNA at -30 dBm microwave power. The false-color represents the S${21}$ transmission parameter. (b) Transmission parameter S${21}$ amplitudes at high (+20 dBm) and low (-30 dBm) microwave powers, measured at external fields of 1.215 kOe and 1.220 kOe, respectively. The frequency splitting between the two peaks corresponds to twice the coupling strength $g$. (c) Microwave power transmission $P_{out}/P_{in}$ (where S${21}$ = 10 $log$($P_{out}/P_{in}$)) as a function of frequency at different microwave powers and magnetic field fixed at 1.1 kOe, far way from the coupling region. (d) Microwave power transmission $P_{out}/P_{in}$ of an unloaded SC resonator as a function of frequency at different microwave powers without applying a magnetic field.
}
\label{Fig3}
\end{figure}

To understand this unexpected observation, we conducted microwave power-dependent VNA measurements, see Fig.~\ref{Fig3}. In Fig.~\ref{Fig3}(a), magnon-photon level repulsion is observed at an input power of $-30$ dBm, where the resonator mode appears significantly stronger compared to the high-power measurement shown in Fig.~\ref{Fig2}(d). As the magnetic field decreases, the magnon mode shifts closer in frequency to the SC resonator mode while its intensity increases.
As shown in SM Fig.~2(a), the frequency separation between the modes becomes minimal, and a clear exchange in their intensities occurs at around $1.22$~kOe. Below this field, the modes diverge again, and the magnon mode weakens as its frequency moves away from the resonator frequency. The relative intensities of the hybridized modes at $-30$~dBm and $+20$~dBm are shown in Fig.~\ref{Fig3}(b). At higher power, the hybridization field (where a minimal frequency separation between upper and lower branches is observed) shifts slightly, to $1.215$~kOe, and the separation between the magnon and resonator modes narrows -- the coupling strength decreases.

Microwave power transmission $P_\mathrm{out}/P_\mathrm{in}$ [where S${21}$ = 10 $log$($P_\mathrm{out}/P_\mathrm{in}$)] spectra as a function of frequency at a fixed magnetic field of $1.1$~kOe for various microwave powers are displayed in Fig.~\ref{Fig3}(c). The resonator mode intensity gradually decreases as microwave power increases, eventually deviating from a Lorentzian lineshape: the signal becomes highly asymmetric, exhibiting an abrupt jump in intensity as the frequency is varied. This behavior is particularly visible at high microwave powers and a signature of mode bifurcation. To gain insights into the observed behavior, the microwave-power dependence of the unloaded resonator was studied. Figure~\ref{Fig3}(d) shows the microwave power  $P_\mathrm{out}/P_\mathrm{in}$ transmission spectra of the unloaded resonator without any magnetic field applied for varying microwave powers. A clear mode bifurcation occurs above approximately $-15$~dBm power due to nonlinearity in the resonator, which likely is the reason for the power-dependent behavior of the resonator loaded with the sample seen in Fig.~\ref{Fig3}(c). We attribute this nonlinearity in the resonator to the local heating of weak links forming at the boundaries of the superconducting NbN grains~\cite{Adbo_PhysRevB_2006}.

A comparison between Figs.~\ref{Fig3}(c) and (d) reveals that loading the resonator with the YIG/Pt sample and applying a magnetic field leads to a substantial decrease in the resonator's quality factor -- from 1210 to 105 -- and a slight downshift in the resonance frequency from approximately 6.03 GHz to 5.9 GHz. Additionally, the resonance lineshape at high power deviates  from an ideal Lorentzian profile. Although mode bifurcation is observed in both cases, the threshold microwave power required to induce nonlinearity differs significantly: for the unloaded resonator, nonlinearity emerges around $-15$~dBm, whereas for the loaded resonator, the onset of nonlinearity is significantly shifted to higher powers, appearing above $+15$ dBm. This can be understood by examining the resonator linewidth: {After the YIG/Pt sample is placed onto the resonator, a noticeable resonator linewidth broadening is observed, which is primarily attributed to microwave losses in the GGG substrate at cryogenic temperatures~\cite{Liu_PhysRevB_2018}. Along with the enhanced linewidth, the intensity of the resonator mode decreases significantly, as shown in Fig.~\ref{Fig3}(c).}

\begin{figure}[t]
\centering
\includegraphics[width=0.99\columnwidth]{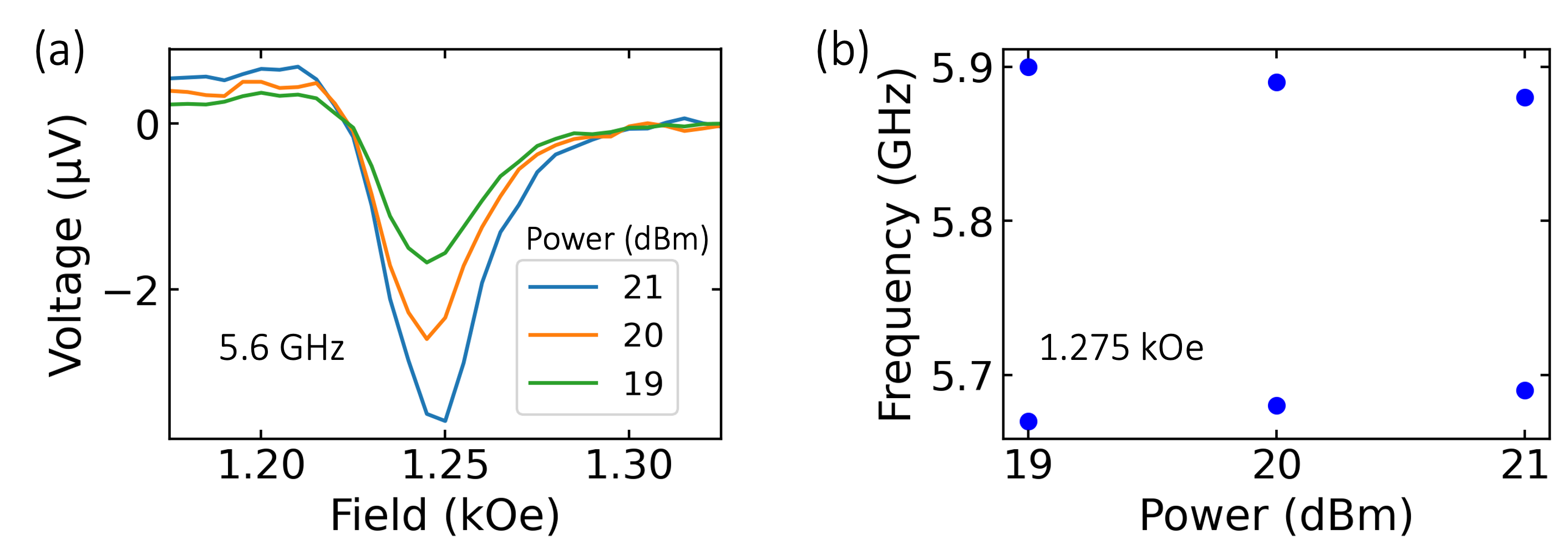}
\caption{(a) Measured spin pumping voltage as a function of the external field at a microwave frequency of 5.6 GHz for three different microwave powers. (b) Mode separation between upper and lower branches at various microwave powers at the hybridization field of 1.275 kOe detected by spin pumping and inverse spin Hall effect.
}
\label{Fig4}
\end{figure}

To corroborate this interpretation further, we performed microwave power-dependent electrical measurements (identical measurement configuration as in Fig.~\ref{Fig2}). As is shown in Fig.~\ref{Fig4}(a), the intensity of the Kittel-like mode -- detected by spin pumping and inverse spin Hall effect -- increases with increasing microwave power at a microwave frequency of 5.6 GHz. 
This indicates that the spin precession amplitude increases with higher microwave power. The resonance frequencies of the Kittel-like mode at the hybridization field of $1.275$~kOe (in the coupling regime, where a gap separates the upper and lower frequency modes) as a function of the applied microwave power
is presented in Fig.~\ref{Fig4}(b) (note that the hybridization field value for these microwave powers remains constant). As the microwave power increases, the frequency separation of the coupled modes gradually decreases. This reduction in peak separation indicates a weakening of the magnon-photon coupling strength with increasing microwave power. This trend is also evident in the VNA measurements, as shown in Fig.~\ref{Fig3}(b). 
Our electrical and microwave spectroscopy measurements suggests that higher microwave powers -- above a certain threshold -- may lead to nonlinear effects in the superconducting resonator
which can suppress the coherent coupling between the magnon and photon modes. 

Furthermore, with the help of the microwave-power dependent experiments (Figs.~\ref{Fig3} and \ref{Fig4}), we can better understand the vanishingly small resonator signal away from the coupling regime at a high microwave power, shown in Fig.~\ref{Fig2}(d): The SC resonator transitions into a highly nonlinear state at sufficiently high microwave powers, which broadens the resonance linewidth and dramatically reduces its intensity (both, in case of the loaded and unloaded system, Figs.~\ref{Fig3}(c) and (d), respectively). Consequently, no signal is observed away from the coupling region, where we would usually expect the highest signal strength (strongest transmission). However, if the magnetic field is tuned so that SC resonator and the Kittel-like modes are degenerate and a mode hybridization occurs, energy from the SC resonator is efficiently transduced into the magnonic system. This leads to two effects: (1) The SC resonator experiences additional losses, affecting both linewidth and signal strength and, hence, destroying the nonlinear state. As a result, the photonic character of the hybridized state may be observable again. (2) The magnonic character of the hybridized state is observed as the Kittel-like mode is sufficiently close to SC resonance frequency. Combined, these two effects explain why two frequency-split modes with relatively large intensity are observed above and below the SC resonator frequency for the high-power experiments shown in Fig.~\ref{Fig2}(b).

\section{Conclusion}
In conclusion, we demonstrated the coherent coupling of magnons with microwave photons, detected electrically via spin pumping and the inverse spin Hall effect, in a superconducting resonator/YIG film hybrid system at cryogenic temperatures. The electrical detection results are in good agreement with those obtained from microwave transmission measurements. As the magnon mode approaches the coupling region, its linewidth increases, indicating enhanced damping. Power-dependent measurements reveal that the coupling strength decreases with increasing microwave power. Additionally, we observed a suppression of the resonator mode at higher microwave powers, which we attribute to nonlinear effects in the SC resonator above a critical threshold microwave power.

\section{Acknowledgement}
We acknowledge support by the U.S. Department of Energy, Office of Basic Energy Sciences, Division of Materials Sciences and Engineering under Award DE-SC0020308. MBJ acknowledges the JSPS Invitational Fellowship for Researcher in Japan.

\bibliography{apssamp}

\end{document}